\documentclass[a4paper,11pt]{article}
\usepackage{pos}
\usepackage{xspace}

\providecommand{\alpS}{\ensuremath{\alpha_\mathrm{s}}\xspace}
\providecommand{\mZ}{\ensuremath{M_\mathrm{Z}}\xspace}
\providecommand{\ttbar}{\ensuremath{\mathrm{t}\bar{\mathrm{t}}}\xspace}

\title{Determination of the strong coupling from jet measurements at CMS}

\author*[a]{Patrick L.S. \textsc{Connor}\note{on behalf of the CMS Collaboration}}

\affiliation[a]{Center for Data and Computing in Natural Sciences,\\
    Universität Hamburg, Germany}

\emailAdd{patrick.connor@desy.de}

\abstract{
     A vast program of determinations of the strong coupling \alpS is being undertaken by CMS. These measurements exploit several QCD dominated processes that are sensitive to \alpS, and present different theoretical and experimental challenges. A review of the current public results and perspective is given. 
}

\FullConference{42nd International Conference on High Energy Physics (ICHEP2024)\\
18-24 July 2024\\
Prague, Czech Republic\\}

\begin{document}
\maketitle

\section{Introduction}

\begin{figure}[h]
    \centering
    \includegraphics[width=0.6\textwidth]{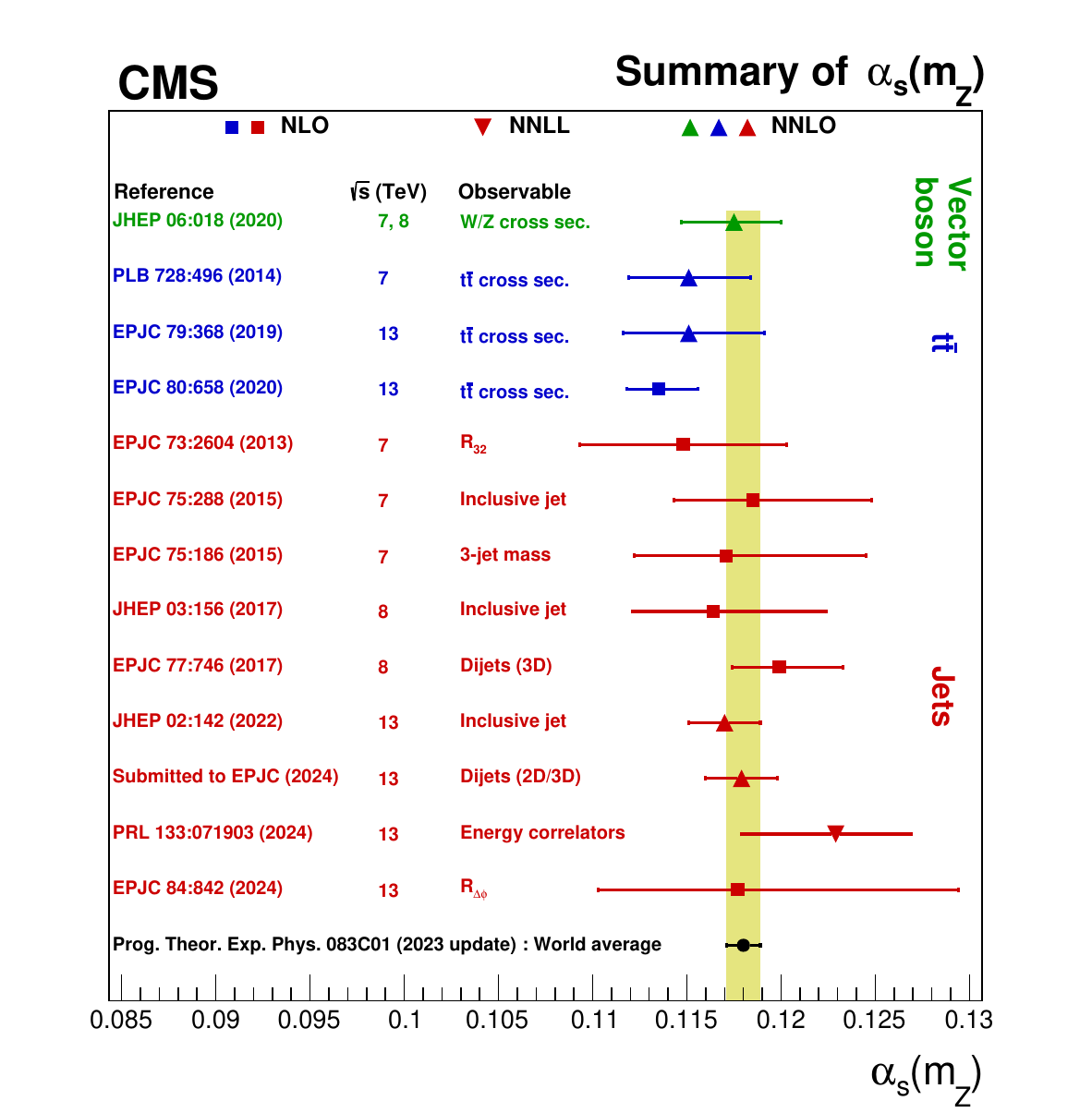}
    \caption{
        Overview of $\alpS(\mZ)$ determinations at CMS.
        The different marker styles denote the accuracy of the fixed-order prediction.
        The various colours indicate the channel.
        The horizontal bars represent the total uncertainty.
        The PDG average is shown in back, and its uncertainty is shown by its own error bar as well as by the shaded band.
    }
    \label{fig:overview}
\end{figure}

Since the beginning of data taking at the CERN Large Hadron Collider (LHC), the CMS Collaboration has been very active in providing determinations of the strong coupling \alpS from measurements of various topologies.
This is illustrated in Fig.~\ref{fig:overview}.
Given their intrinsic nature, jet measurements are obviously a fecund proxy to precise determinations of~\alpS.
In these proceedings, we shall review the determinations of the strong coupling constant $\alpS(\mZ)$ from jet measurements with CMS proton data recorded in LHC Run~1 and Run~2.
Other channels as well as the running of the strong coupling are not discussed here.

To the exception of the extraction of \alpS from jet substructure, which we will describe last, all determinations rely on the factorisation of proton-proton collisions~\cite{Collins:1989gx}:
\begin{align} \label{eq:factorisation}
    \sigma_{pp} &= \sum_{ij\in{gq\bar{q}}} f_i(x_i,\mu_F^2) \otimes f_j(x_j,\mu_F^2) \otimes \hat{\sigma}_{ij} \left(x_i,x_j,\frac{Q^2}{\mu_F^2},\frac{Q^2}{\mu_R^2},\alpS(\mu_R^2)\right)
\end{align} 
where $\sigma_{pp}$ corresponds to the hadronic cross section, which may be obtained from the experimental data or calculated from the right-hand side.
The summation symbol runs over all flavours of quarks, antiquarks, and gluons.
The $f$ symbols denote the parton distribution functions~(PDFs), which describe the contributions of the protons, as opposed to the partonic cross section $\hat{\sigma}$, which describes the hard part of the interaction and may be obtained from fixed-order (FO) calculations.
A factorisation scale $\mu_F$ is used to separate the respective contributions.
In general, collinear PDFs are assumed, in which case the momentum fraction $x$ is sufficient to parameterise the parton kinematics.
The $\hat{\sigma}$ exhibits the dependence in \alpS, which most measurements exploit, but \alpS also intervenes in the PDF evolution.
The \alpS itself depends on a renormalisation scale~$\mu_R$, which is most of the time taken as =$\mu_F$.

The simplest approach (method~\#1) consists in taking existing PDF sets provided for different value of \alpS, calculate $\sigma_{pp}$ accordingly, extract a $\chi^2$ from the comparison with some experimental observable, and deduce the value of \alpS that best describes the data.
This approach is technically easier, however it does not handle correctly the correlations between the \alpS in the PDF evolution and the \alpS in $\hat{\sigma}$.
To mitigate this effect, one possibility consists in fitting cross section ratios, where the impact of PDFs is reduced (although it certainly does not vanish completely).
A more advanced approach (method~\#2) consists in extracting PDFs and \alpS in a combined fit procedure.
At CMS, this has so far always been done with the xFitter software~\cite{Bertone:2017tig}.
Such a procedure is more complex to set up but yields more precise and more accurate results at the same time.
Note that Eq.~\ref{eq:factorisation} does not include non-perturbative~(NP) corrections, which cover for hadronisation and multi-parton interactions.
These effects are however accounted for in the extractions of \alpS exposed in the next section (both methods).

Finally, it is possible to extract \alpS from the jet substructure.
This approach does not rely at all on Eq.~\ref{eq:factorisation} but on the branchings happening within the jet.
Novel observables such as the so-called energy correlators~\cite{PhysRevD.102.054012} permit an extraction of \alpS too with different properties of the same events.

\section{Review}

Historically, the first extraction of \alpS with CMS was performed on data recorded at 7~TeV with the $R_{32}$ observable~\cite{CMS-PAPERS-QCD-11-003}:
\begin{equation*}
    R_{32} = N^\text{eff}_\text{incl. 3-jet} / N^\text{eff}_\text{incl. 2-jet} \sim \alpS
\end{equation*}
The data were fit with NLO predictions following method~\#1:
\begin{equation}
    \alpS(\mZ) = 0.1148\pm0.0014(\text{fit})\pm0.0018(\text{PDF})\pm0.0050(\text{theory})
\end{equation}
where the theory uncertainties by far dominate the value.

Then the inclusive jet double differential cross section has been measured at various centre-of-mass (c.m.s.) energies~\cite{CMS:2012ftr,CMS:2016lna,CMS:2021yzl}.
The data were fit with the state-of-the-art FO predictions available at the time of the respective publications:
the extractions at 7 and 8~TeV were performed with NLO predictions and were dominated by scale uncertainties, whereas the 13~TeV measurement was fit with NNLO interpolation tables~\cite{Britzger:2022lbf,Britzger:2012bs} following method~\#2.
The latter has yielded to date the most precise \alpS with CMS data:
\begin{equation}
    \alpS(\mZ) = 0.1166 \pm 0.0014~(\text{fit}) \pm 0.0007~(\text{model}) \pm 0.0004~(\text{scale})\pm 0.0001~(\text{param.})
\end{equation}
where the fit uncertainties now dominate.

\begin{figure}[h]
    \centering
    \includegraphics[width=\textwidth]{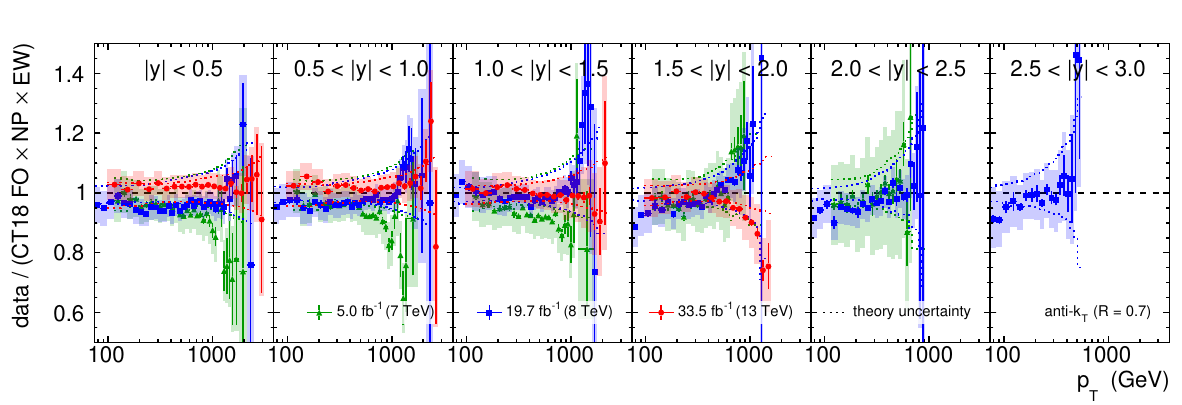}
    \caption{
        Overview of inclusive jet measurements.
        Each cell corresponds to a rapidity interval.
        The $x$-axis corresponds to the jet transverse momentum, whereas the $y$-axis represents the ratio of experimental data points with fixed-order predictions at NNLO accuracy obtained with CT18 PDFs and corrected with non-perturbative and virtual electroweak corrections.
        Each c.m.s.~energy is shown with a different colour.
        The shaded bands (dotted lines) indicate the experimental (theoretical) uncertainty.
    }
    \label{fig:incljet}
\end{figure}

Recently, NNLO interpolation tables have been made available at lower c.m.s.~energies.
A comparison of data and predictions at 7, 8, and 13~TeV is shown in Fig.~\ref{fig:incljet}.
In general, they agree within uncertainties, although the absolute normalisation seem to differ and although the respective ratios exhibit different shapes.
This may happen for various reasons.
First, the non-perturbative corrections are still taken from the respective publications and are not exactly consistent.
Then, the unfolding procedure has significantly changed over the years, as well as the jet energy calibration.
Note that even if those had been consistent, a direct comparison such as the one on Fig.~\ref{fig:incljet} should still taken with a pinch of salt, as the various measurements are also partly systematically correlated.

The CMS Collaboration has also released several dijet cross section measurements at 7, 8, and 13 TeV in different versions: there are double and triple differential cross sections as a function of the average transverse momentum or of the mass, and as a function of different combinations of the rapidity of the two leading jets~\cite{CMS:2012ftr,CMS:2017jfq,CMS:2023fix}.
The choice of the observables has been guided over the years by discussions with theorists to find the one with the highest sensitivity to~\alpS and to PDFs.
They typically yield a similar precision as inclusive jet cross sections.
To be noted, certain experimental effects reduce significantly, such as the jet energy resolution.

A trijet mass double differential cross section was also performed with 7~TeV data~\cite{CMS-PAPERS-SMP-12-027} but was not repeated at higher c.m.s.~energies.
This observable, proportional to $\alpS^2$, also exhibits interestingly low uncertainties.
In general, using method~\#1, it led to small uncertainties than the analog analysis with inclusive jet cross section to the exception of scale uncertainties.
Once a better prescription to handle scale uncertainties is found, it will be interesting to return to this observable.

To conclude the list of extractions of \alpS based on Eq.~\ref{eq:factorisation}, we want to mention the novel observable $R_{\Delta\phi}$, which is a sophisticated type of ratio observables~\cite{CMS:2024hwr}.
It exhibits very small experimental uncertainties, taking advantage of their cancellation in the ratio.
However, since FO predictions are only available at NLO to date, the resulting \alpS is not yet competitive.
We look forward to getting NNLO predictions from the theory community.

Finally, the CMS Collaboration has also recently measured a ratio observable of three- and two-particle energy correlators using 13~TeV data~\cite{CMS:2024mlf}.
This observable was fit with aNNLL predictions, and yielded the most precise extraction of \alpS from jet substructure:
\begin{equation}
    \alpS(\mZ) = 0.1229 \pm 0.0014~(\text{fit}) {}^{+0.0014}_{-0.0012}(\text{stat}) {}^{+0.0023}_{-0.0036}(\text{syst})
\end{equation}

\section{Discussion}

An overview table of all determinations of \alpS by the CMS Collaboration with a detailed breakdown of the uncertainties may be found in Table~6 of Ref.~\cite{CMS:2024gzs}.
In general, no tension has been observed with the PDG average.
However, a direct comparison of the various measurements is a dangerous exercise, as the different analyses may have undergone slightly different procedures and calibrations, and as uncertainties may have slightly different meanings despite the same name.
Furthermore, measurements obtained from the same statisical data set have large statistical correlations; similarly, measurements of the same observable have large systematic correlations.
In both cases, these correlations have not been released by CMS.

Therefore, the greatest care must be taken when comparing directly with one another.
In the following paragraph, we allow ourselves certain careful observations.
First, we observe that ratio observables have smaller uncertainties than differential cross sections.
This emphasises the relevance of such observables.
Second, model uncertainties are not always provided, and when they are, they are derived following different procedures.
Similarly, the non-perturbative (NP) corrections and uncertainties seem to matter, but are derived following the state-of-the-art procedure known at the time of the respective publications.
Both model and NP uncertainties are not Gaussian, but, in the absence of a clear prescription to handle them, are treated as such in the QCD interpretation.
Third, we find that determinations at NNLO are dominated by the fit uncertainties.
These fit uncertainties are mostly (but not exclusively) arising from the experimental uncertainties.

Given this last point, the CMS Collaboration is considering several complementary approaches to improve the precision further: first, explore new observables such as novel cross section ratios.
The $R_{\Delta\phi}$ is a good example of a novel observable with good sensitivity.
Second, existing measurements in various channels (vector boson, jet, and \ttbar) may be combined, as well as analog measurements at various c.m.s.~energies.
The former approach has already been applied in Ref.~\cite{CMS:2021yzl}, but the latter has not yet been released by the Collaboration.
Third, improve the calibration, in particular the jet calibration.
Finally, the CMS Collaboration is investigating the simultaneous measurement a several observables.
This way, we may take advantage of the advantages of the respective observables: for instance, the cross section ratios have enhanced sensitivity to \alpS with respect to absolute cross sections, but the information from the absolute normalisation is lost.
Another advantage would be to combine inclusive jet and dijet measurements, as systematic effects propagate differently.
This approach is discussed in details in Ref.~\cite{Collaboration:2888301}.

\section{Conclusions}

The CMS Collaboration has provided numerous determinations of the strong coupling.
All are found in agreement with the PDG value.
With the advent of predictions at NNLO, the fit uncertainty has become dominant.
Prospects have been discussed, such as the exploration of new observables, the combination of existing measurements at different centre-of-mass energies and in different channels, the refinement of the event reconstruction, and the simultaneous determination of statistically correlated observables.

\small

\bibliographystyle{JHEP}
\bibliography{main}

\end{document}